\documentclass[a4paper,11pt]{article}
\usepackage{jheppub} 
\usepackage[T1]{fontenc} 
\usepackage{hyperref}
\usepackage{latexsym}
\usepackage{graphics}
\usepackage{graphicx}
\usepackage{feynmf}
\usepackage{color}

\newcommand{\be}{\begin{equation}}
\newcommand{\ee}{\end{equation}}
\newcommand{\ba}{\begin{eqnarray}}
\newcommand{\ea}{\end{eqnarray}}
\newcommand{\ar}{\arrowvert}

\newcommand{\tr}[1]{\textrm{#1}}
\newcommand{\bw}{\begin{widetext}}
\newcommand{\ew}{\end{widetext}}

\newcommand{\Imag}{\mathop{\mathrm{Im}}}

\begin{document}
\title{\boldmath One-loop $W_LW_L$ and  $Z_LZ_L$ scattering from the Electroweak Chiral Lagrangian 
with a light Higgs-like scalar}
\author{Rafael L. Delgado, Antonio Dobado and Felipe J. Llanes-Estrada}
\affiliation{Departamento de F\'isica Te\'orica I, Universidad Complutense de Madrid, 28040 Madrid, Spain}
\abstract{
By including the recently discovered  Higgs-like scalar $\varphi$ in the Electroweak Chiral Lagrangian, 
and using the Equivalence Theorem, we carry out the complete one-loop computation of the elastic scattering amplitude for the longitudinal components of the gauge bosons $V=W, Z$ at high energy. We also compute $\varphi\varphi \rightarrow \varphi\varphi$ and the inelastic process $VV\rightarrow \varphi\varphi$, 
and identify the counterterms needed to cancel the divergences, namely the well known $a_4$ and $a_5$ chiral parameters plus three additional ones only superficially treated in the literature because of their dimension 8.
Finally we compute all the partial waves and discuss the limitations of the one-loop computation due to only approximate unitarity.
}
\maketitle
\section{Introduction}
The LHC directly probes for the first time the sector of the Standard Model responsible for Electroweak Symmetry Breaking.
Two-particle invariant mass spectra
of the longitudinal components of gauge boson pairs $W_L W_L$  and $Z_L Z_L$ are not yet at hand, but expected in the next years. As the remainder of the Goldstone bosons of 
electroweak symmetry breaking, the scattering of the longitudinal bosons at
high-energy (high compared with $M_W$, but not larger than about $4\pi v\simeq 3\,{\rm TeV}$)  is predicted by theory through the equivalence theorem~\cite{ET}, even in the presence of strong interactions that may make other predictions doubtful.

The finding that the LHC collaborations ATLAS~\cite{ATLAS} and CMS~\cite{CMS} have published~\cite{twophotons}
is a boson with scalar quantum numbers and couplings compatible with those of a Standard Model Higgs. This might bring the Minimal Standard Model (MSM) to closure.

Most interestingly, no further new particle has been sighted~\cite{searches} in the first run of the LHC, up to an energy of 600-700 GeV 
(and higher yet for additional vector bosons). 
This  mass gap in the spectrum also
naturally suggests that the Higgs is an additional Goldstone boson, perhaps a dilaton from spontaneous breaking of scale invariance, or from a composite Higgs model based on $SO(5)/SO(4)$ or any other coset. The effective Lagrangian approach includes these cases as we will point out, but irrespective of the true nature of the scalar boson, it encodes its interactions with the rest of the symmetry  breaking sector. One feature that we will adopt from these models, though, also shared by the Standard Model, is that the Higgs-potential self-couplings are of order $M_\varphi^2$, and thus negligible for $s\gg M_\varphi^2$. Apart from this assumption, that covers all models of interest at the present time, our discussion will remain general and uncommitted to a particular new physics scenario.

Several groups\cite{Espriu:2013fia,Azatov:2012bz,Brivio:2013pma,Alonso:2012px,Pich:2013fba,Jenkins:2013zja,Degrande:2012wf,Buchalla:2013rka,Buchalla:2012qq}  are studying in detail the formulation of effective Lagrangians for the four visible particles, their scattering amplitudes at low-energy and the unitarization of those amplitudes to reach higher energies. These studies extend traditional effective-theory approaches~\cite{Appelquist} to the Electroweak Symmetry Breaking Sector (EWSBS) modeled in total analogy to Chiral Perturbation Theory in QCD~\cite{ChPT}.

In a recent work~\cite{Delgado:2013loa} we have shown that, for essentially any parameter choice except that of the Standard Model and perhaps other very carefully tuned sets, the interactions will generically become strong at sufficiently high energy, and have argued that a second, very broad scalar pole is expected.

In this article we complete the one-loop computation of the two-body scattering amplitudes among the $\omega$ Goldstone bosons and the $\varphi$ scalar with such a generic effective Lagrangian, in the kinematic regime
$M_\varphi^2\ll s<4\pi v\simeq 3\,{\rm TeV}$. The Lagrangian density is exposed in
section~\ref{sec:eff_Lag} and the scattering amplitudes derived therefrom, in dimensional regularization, are given in section~\ref{sec:amplitudes}. The calculation has been performed both analytically and also with standard one-loop automated computer tools, and the results agree. The Feynman diagrams resulting from the effective Lagrangian are delayed to the appendix given their large number.

Renormalization is carried out in section~\ref{sec:renorm}. Five NLO coefficients are necessary, the well known $a_4$ and $a_5$ from the Electroweak Chiral Lagrangian, and three less studied ones, also multiplying operators of dimension 8, that renormalize the Higgs self-interactions and the channel coupling between scalar and longitudinal vector bosons. 
We do not assess operators that are unnecessary to carry out the renormalization of the one-loop computation, 
with the exception of $(\varphi\partial_\mu \varphi) (\varphi\partial^\mu \varphi)$ that we examine in subsection~\ref{subsec:tree}; 
the interested reader can find a table of the 59 dimension-6 operators that extend the SM in~\cite{Jenkins:2013zja}.

In section~\ref{sec:partialwaves} we provide the partial-wave projections of all three two-body amplitudes, that will prove necessary in future work to examine the possible existence of new resonances channel by channel.

Section~\ref{sec:numeric} shows a numerical computation of the various partial waves to gain a feeling for their behavior and sensitivity to the unknown parameters that carry the theory beyond the Standard Model, and also to expose the violation of unitarity in perturbation theory, since in the effective Lagrangian approach, amplitudes grow like a power of Mandelstam-$s$. 

Our findings are summarized in section~\ref{sec:summary}.

\newpage

\section{The Electroweak Chiral Lagrangian with a Light Scalar} \label{sec:eff_Lag}

One of the lowest-order equivalent forms of the universal Electroweak Chiral Lagrangian with the known particle content is a gauged $SU(2)_L \times SU(2)_R/SU(2)_C = SU(2) \simeq S^3$ Non-linear Sigma Model (NLSM) coupled to a scalar field $\varphi$ as
\be 
\label{genericLagrangian}
{\cal L}_0=\frac{v^2}{4}g(\varphi /f)\tr(D_\mu U)^\dag
D^\mu U+\frac{1}{2}\partial_\mu \varphi \partial^\mu
\varphi-V(\varphi) 
\ee 
where $U$ is a field taking values in the $SU(2)$ coset that can be parametrized for example as $U=\sqrt{1-\tilde
\omega^2/v^2}+ i \tilde  \omega/v$; $\tilde \omega =
\omega_a\tau^a$ being the would-be Goldstone boson (WBGB) field~\cite{GB}. The $SU(2)_L\times U(1)_Y$ subgroup is gauged as usual through the covariant derivatives
$D_\mu U=\partial_\mu U + W_\mu U- U  Y_\mu$,
$W_\mu = - g i W_\mu^i \tau^i/2$, $Y_\mu = - g' i B_\mu^i \tau^3/2$. In terms of Fermi's weak constant, $v^2:=1/(\sqrt{2}G_F)=(246\,{\rm GeV})^2$, while $f$ is an arbitrary, new-physics  energy scale 
controlling the generic dynamics of the EWSBS. The scalar field interacts through  $g(x)$, an arbitrary analytical functional; in the effective-theory approach only the first terms of its Taylor expansion are probed  
\be \label{gexpansion}
g(\varphi /f)= 1 +2 \alpha \frac{\varphi}{f} +\beta\left(\frac{\varphi}{f}\right)^2+..
\ee 
Here we have introduced two parameters $\alpha$ and $\beta$ instead of the more common $a$ and $b$ in~\cite{scalar}, but clearly we have $a= \alpha v/f$ and
$b= \beta v^ 2/ f^ 2$. 
With this natural but maybe unconventional choice, having $f$ instead of $v$ in the denominators,
the value of the adimensional vacuum-tilt parameter $\xi \equiv v^2/ f^2$ that corresponds to the Standard Model is $\xi=1$. 

Our philosophy here is to weigh the WBGB field intensity against the EWSB scale $v$ and the scalar field 
$\varphi$ against the possible new scale $f$. 
Of course $f=v$ is a particular possibility corresponding to $\alpha=a$ and $\beta = b$  (see \cite{bounds} for  some recent experimental bounds on the $a$ and $b$ that we have also briefly discussed in~\cite{Delgado:2013loa}). Finally 
$V$ is an arbitrary analytical potential for the scalar field,
\ba \label{potential}
V(\varphi) =  \sum _{n=0}^{\infty}V_n \varphi^n
\equiv  V_0 + 
\frac{M_\varphi^2}{2} \varphi^2 +\lambda_3 \varphi^3+\lambda_4 \varphi^4+...
\ea
At the the next to leading order in the chiral expansion one should add the four derivative terms
\begin{eqnarray} \label{higherorderL}
{\cal L}_4 & = &  a_4(tr V_\mu V_\nu)^2 +  a_5(tr V_\mu V^\mu)^2 \\ \nonumber
 & + &\frac{\gamma}{f^4} (\partial_\mu\varphi \partial^\mu\varphi)^2
 +  \frac{\delta}{f^2} (\partial_\mu\varphi \partial^\mu\varphi)tr(D_\nu U)^\dag D^\nu U
+\frac{\eta}{f^2} (\partial_\mu\varphi \partial^\nu\varphi)tr(D^\mu U)^\dag D_\nu U+...
\end{eqnarray}
where $V_\mu= D_\mu U U^\dagger$. We have written explicitly only the five terms strictly needed for the renormalization of the one-loop elastic WBGB scattering amplitudes (for $s\gg M_W^2$) and the unitarity-related processes  $\omega\omega \rightarrow \varphi\varphi$ and $\varphi\varphi  \rightarrow \varphi\varphi$.
These terms produce additional contributions to the amplitudes which are of order $s^2$. 

The chiral parameters $a_4$ and $a_5$ (multiplying the operators $O_{D1}$ and $O_{D2}$ in the classification of~\cite{Buchalla:2013rka})
and the new ones $\gamma$, $\delta$ and $\eta$  
depend on  whatever unknown underlying dynamics responsible for the spontaneous symmetry breaking of electroweak interactions might exist. They all vanish in the MSM.
The operators with coefficient $\delta$ and $\eta$ are identified as $O_1$ and $O_2$ in the classification of Azatov {\it et al.}~\cite{Azatov:2012bz} while they are $P_{19}$ and $P_{20}$ in~\cite{Brivio:2013pma} and are NLO equivalent to $O_{D7}$, $O_{D8}$ in~\cite{Buchalla:2013rka}. The operator associated with $\gamma$ is denoted as $P_H$ in~\cite{Brivio:2013pma} and $O_{D11}$ in~\cite{Buchalla:2013rka}. None of these authors give much detail on the use or scale-dependence of these operators, important to this work.

The two operators multiplying $\delta$ and $\eta$ are apparently of dimension 6. But this leading dimension affects only transverse gauge-boson inelastic scattering $W_TW_T\to\varphi\varphi$ and not the longitudinal ones. When expanding $U$, the relevant $\omega\omega\to \varphi\varphi$ terms
are of dimension 8 as shown shortly in Eq.~(\ref{bosonLagrangian}). Thus, they are apparently of a high order in the classification of all operators beyond the Standard Model, but as we will see they are necessary already in one-loop renormalization.

 The Lagrangian in Eq.~(\ref{genericLagrangian}) and Eq.~(\ref{higherorderL}) is able to reproduce the low-energy physics of this sector of the SM for any possible dynamics having at least an approximate $SU(2)$ custodial isospin symmetry in the limit $g=g'=0$. For example the MSM corresponds to the parameter selection $\alpha=\beta=\xi=1$ and $a_4=a_5=\gamma=\delta=\eta=0$. The Higgs field $H$ is just the scalar field  $\varphi$ so that $M_H^2=M_\varphi^2= 2\lambda v^2$, and the scalar self-couplings are $\lambda_3= \lambda v$, $\lambda_4 = \lambda/4$ (both proportional to $M_\varphi^2$) and $\lambda_i=0$ for $i \ge 4$. 

In dilaton models~\cite{Grinstein} $\varphi$ would represent the dilaton field, 
$\alpha=\beta = 1$ as in the MSM but $\xi$ is arbitrary, $f$ being the scale of the symmetry breaking. The potential and NLO parameters depend on the particular dilaton model but in any case $\lambda_i$ is of order $M^2_\varphi$ for any $i$. 

Third, we also have the example of the $SO(5)/SO(4)$ Minimally Composite Higgs Model~\cite{SO(5)} where $\alpha= \cos \theta /\sqrt{\xi}$, $\beta = \cos(2 \theta)/\xi$, $\sin \theta =\sqrt{\xi}$ and $a_4$, $a_5$ and the scalar-boson couplings depend on the particular details of the model, but it can be assumed that the $\lambda_i$ are of order  $M_\varphi^2$ too.

Finally it is also possible to reproduce the old Higgsless Electroweak Chiral Lagrangian (EWChL) in \cite{Appelquist}  by the simple parameter choice $\alpha= \beta =\gamma=\delta=\eta=0$.

As discussed in the introduction, we pursue the elastic scattering of the longitudinal components of the electroweak bosons at high energies, i. e., for $\sqrt{s}\gg 100\,{\rm GeV}$. In this case, we can apply the Equivalence Theorem:
\be
T(\omega^a\omega^b \rightarrow \omega^c\omega^d )  
= T(W_L^aW_L^b \rightarrow W_L^cW_L^d ) + O\left(\frac{M_W}{\sqrt{s}}\right)\ , 
\ee 
and thus we will be probing the WBGB dynamics. This theorem applies for any renormalizable gauge, but the Landau gauge (where there remain massless WBGB) turns out to be particularly useful. Therefore, in the following we will set $g=g'=0$ and the only degrees of freedom to be considered will be the massless (in the Landau gauge) WBGB and the Higgs-like scalar $\varphi$. Moreover, according to ATLAS and CMS $M_\varphi \simeq 125$ GeV.   Then $M_\varphi \sim M_W \sim M_Z \sim 100\,{\rm GeV}$.
As a consequence it is a perfectly consistent approximation to consider the massless $\varphi$ limit, i.e., $M_\varphi \simeq 0$ if one is only interested in the energy region where the ET can be applied. Therefore we will concentrate on the WBGB scattering for  $M^2_\varphi, M^2_W, M^2_Z \simeq 0\ll s < \Lambda^2$ where $\Lambda$ is some ultraviolet (UV) cutoff of about 3 TeV, setting the limits of applicability of the effective theory. 

As in the three particular models just mentioned, we will also assume that the $\lambda_i$ pure-scalar potential parameters are of order $M_\varphi^2$ so that we can neglect the scalar potential altogether. 
In this kinematic regime, the relevant Lagrangian, derived  from Eqs.~(\ref{genericLagrangian}) through~(\ref{higherorderL}) is
\ba \label{bosonLagrangian}
{\cal L} & = & \frac{1}{2}\left(1 +2 \alpha \frac{\varphi}{f} +\beta\left(\frac{\varphi}{f}\right)^2\right)
\partial_\mu \omega^a
\partial^\mu \omega^b\left(\delta_{ab}+\frac{\omega^a\omega^b}{v^2}\right)   
\nonumber +\frac{1}{2}\partial_\mu \varphi \partial^\mu \varphi  \nonumber  \\
 & + & \frac{4 a_4}{v^4}\partial_\mu \omega^a\partial_\nu \omega^a\partial^\mu \omega^b\partial^\nu \omega^b +
\frac{4 a_5}{v^4}\partial_\mu \omega^a\partial^\mu \omega^a\partial_\nu \omega^b\partial^\nu \omega^b  +\frac{\gamma}{f^4} (\partial_\mu\varphi \partial^\mu\varphi)^2  \nonumber   \\
 & + & \frac{2\delta}{v^2f^2} \partial_\mu \varphi \partial^\mu \varphi\partial_\nu \omega^a  \partial^\nu\omega^a
+\frac{2\eta}{v^2f^2} \partial_\mu\varphi \partial^\nu\varphi\partial^\mu \omega^a \partial_\nu\omega^a.
\ea 

Notice also that by rescaling $f$ and redefining $\beta$ it is possible to set $\alpha=1$ in Eq.~(\ref{gexpansion}) without losing generality,
\be \label{gexpansion2}
g(\varphi /f)= 1 +2  \frac{\varphi}{f'} +\beta'\left(\frac{\varphi}{f'}\right)^2+\dots
\ee 
This leaves as free parameters in the above Lagrangian in our energy region of interest the redefined $f$ and $\beta$, the chiral parameters $a_4$ and $a_5$, and the three $\gamma$, $\delta$, $\eta$ ones involving the new scalar boson.  However, in the following we will still keep the explicit $\alpha$-dependence in our formulae so that we can easily  trace for comparison with previous works. In particular, as already pointed out, the old EWChL without any Higgs-like light resonance corresponds to $\alpha=\beta=0$ (and vanishing higher order inelastic couplings). 

\section{The WBGB scattering amplitude in EWChPT at the one-loop level}
\label{sec:amplitudes}
\subsection{Elastic $\omega\omega$ scattering}
In this section we compute the scattering amplitudes using the Landau gauge and  dimensional regularization. We start by elastic WBGB scattering.
Due to the custodial symmetry of the SBS of the SM in the limit $g=g'=0$ the WBGB amplitude $\omega_a\omega_b \rightarrow \omega_c\omega_d$ can be written as
\be
{\mathcal A}_{abcd}= A(s,t,u)\delta_{ab}\delta_{cd}+A(t,s,u)\delta_{ac}\delta_{bd}+A(u,t,s)\delta_{ad}\delta_{bc}
\ee 
because the four particles are identical, the amplitude has to be crossing-symmetric and expressible in terms of only one amplitude $A$.
This, we conveniently expand  following the chiral counting, and also separately quote the NLO tree-level and 1-loop subamplitudes as
\be \label{loopexpansion}
A = A^{(0)} + A^{(1)} \dots =  A^{(0)} + A^{(1)}_{\rm tree} + A^{(1)}_{\rm loop} \dots
\ee

Then, from the Lagrangian in Eq.~(\ref{bosonLagrangian}) the following tree-level amplitude results
\begin{equation} \label{Atree}
A^{(0)}(s,t,u) + A^{(1)}_{\rm tree}(s,t,u) = (1-\alpha^2\xi)\frac{s}{v^2} + \frac{4}{v^4}\left[2a_5 s^2 + a_4(t^2 + u^2)\right].
\end{equation}
At the one-loop level, a lengthy computation of the Feynman diagrams in the appendix gives
\begin{equation} \label{Aloop}
A^{(1)}_{\rm loop}(s,t,u) = \frac{1}{36 (4\pi)^2 v^4}[f(s,t,u)s^2 +(\alpha^2\xi-1)^2( g(s,t,u) t^2 + g(s,u,t) u^2)]
\end{equation}
where we have defined auxiliary functions
\begin{eqnarray}
f(s,t,u) &:=& 
[20 - 40 \alpha^2\xi + \xi^2(56 \alpha^4 - 72 \alpha^2 \beta + 36 \beta^2)] \nonumber\\
& + &%
[12 - 24 \alpha^2\xi + \xi^2(30 \alpha^4- 36 \alpha^2 \beta + 18 \beta^2)] N_\varepsilon\nonumber\\  \label{ffunction}
& + & %
[-18 + 36 \alpha^2\xi +\xi^2(- 36 \alpha^4 + 36 \alpha^2 \beta - 18 \beta^2)] \log\left(\frac{-s}{\mu^2}\right) \nonumber\\
& + & %
3 (\alpha^2\xi-1)^2 \left[\log\left(\frac{-t}{\mu^2}\right) +
\log\left(\frac{-u}{\mu^2}\right)\right] \\ \label{gfunction}
g(s,t,u) &:=& 
26 + 12 N_\varepsilon 
-9 \log\left[-\frac{t}{\mu^2}\right]
-3 \log\left[-\frac{u}{\mu^2}\right]
\end{eqnarray}
and in dimensional regularization $D=4-\epsilon$ the pole is contained as usual in
\be
N_\epsilon =\frac{2}{\epsilon} + \log 4\pi -\gamma\ .
\ee

Because of the factors of $\xi$, the amplitude in Eq.~(\ref{Aloop}) contains terms proportional to $1/v^4$, $1/(v^2f^2)$, and $1/(f^4)$,
reflecting the various possible intermediate states in the one-loop computation.  We have checked also that  our results agree with those found in \cite{Espriu:2013fia} in the limit of vanishing light scalar mass.

\subsection{Scattering amplitudes involving the new $\varphi$ scalar boson}
The next two-body processes to consider are the channel coupling 
$ \omega_a\omega_b  \rightarrow  \varphi \varphi $  
between two $\omega$ WBGB  and a scalar boson pair and $ \varphi \varphi  \rightarrow  \omega_a\omega_b   $, that are needed to obtain one-loop unitarity in $\omega\omega$ scattering. Obviously both processes have the same amplitude because of time reversal invariance. 
Since $\varphi$ is an isospin singlet, the amplitude can be expressed as
\begin{equation}
{\mathcal M}_{ab}(s,t,u) = M(s,t,u) \delta_{ab}.
\end{equation}
Performing the chiral expansion as in Eq.~(\ref{loopexpansion}), we find 
at   tree level, 
\begin{equation} \label{Mtree}
M^{(0)}_{\rm tree}(s,t,u) + M^{(1)}_{\rm tree}(s,t,u)
= (\alpha^2-\beta)\frac{s}{f^2}+ 
\frac{2 \delta}{v^2f^2} s^2+ \frac{\eta}{v^2f^2}(t^2+u^2)
\end{equation}
that takes a one-loop correction:
\begin{equation} \label{Mloop}
M^{(1)}_{\rm loop}(s,t,u) = \frac{\alpha^2-\beta}{576\pi^2 f^2}\left[f'(s,t,u)\frac{s^2}{v^2}  +\frac{\alpha^2 - \beta}{f^2}[ g(s,t,u)t^2 + g(s,u,t)u^2]\right]
\end{equation}
where
\begin{eqnarray}
f'(s,t,u) &=& 
-8 [-9 + \xi (11 \alpha^2 - 2 \beta)]
-6  N_\varepsilon [-6 + \xi(7 \alpha^2 - \beta)]  \\ \nonumber
&  + &
36 (\alpha^2\xi - 1)\log\left[-\frac{s}{\mu^2}\right] +
3 \xi (\alpha^2 - \beta) \left(\log\left[-\frac{t}{\mu^2}\right] + 
\log\left[-\frac{u}{\mu^2}\right]\right) 
\end{eqnarray}
and the function $g$ is as defined in Eq.~(\ref{gfunction}).

Finally we have the amplitude for the elastic scattering $\varphi\varphi\rightarrow\varphi\varphi$, 
\be
{\mathcal T}(s,t,u) =
 T^{(0)} + T^{(1)}_{\rm tree} + T^{(1)}_{\rm loop} \dots
\ee
The tree amplitude is
\be \label{Ttree}
 T^{(0)}(s,t,u) + T^{(1)}_{\rm tree}(s,t,u)=
\frac{2\gamma}{f^4}(s^2+t^2+u^2)
\ee
and the one-loop piece can be written in terms of only one function
\begin{equation}
T(s) = 2 + N_\varepsilon - \log\left(-\frac{s}{\mu^2}\right)
\end{equation}
as
\be \label{Tloop}
T^{(1)}_{\rm loop}(s,t,u) = \frac{3(\alpha^2-\beta)^2}{2(4\pi)^2 f^4}\left[T(s)s^2 + T(t)t^2 + T(u)u^2\right] \ .
\ee

\section{Renormalization of the amplitudes}\label{sec:renorm}
Comparing the tree-level amplitudes in Eqs.~(\ref{Atree}), (\ref{Mtree}), (\ref{Ttree}) with the loop ones in Eqs.~(\ref{Aloop}), (\ref{Mloop}), (\ref{Tloop}) we see that  the divergences in the one-loop pieces can be absorbed just by redefining the couplings $a_4$, $a_5$, $\gamma$, $\delta$ and $\eta$ from the NLO tree-level Lagrangian. Therefore no $\alpha$, $\beta$, $v$, $f$, wave-function nor mass renormalization is needed to obtain a finite amplitude (a pleasant feature of dimensional regularization). 

We proceed by choosing the modified minimal-substraction  or $\overline{MS}$ scheme, so the renormalized couplings are given by
\begin{eqnarray}
a_4^r & = & a_4+ \frac{N_\epsilon}{192\pi^2}(1-\xi \alpha^2)^2   \nonumber \\
a_5^r & = & a_5+\frac{N_\epsilon}{768 \pi^2} (2+5 \xi^2\alpha^4-4 \xi \alpha^2-6\xi^2\alpha^2\beta+3\xi^2\beta^2)\nonumber \\
\gamma^r & = & \gamma+\frac{3N_\epsilon}{64 \pi^2}(\alpha^2-\beta)^2   \nonumber  \\
\delta^r & = & \delta -\frac{N_\epsilon}{192 \pi^2 }(\alpha^2-\beta)(7\xi \alpha^2-\xi\beta-6)  \nonumber  \\
\eta^r  & = & \eta+ \frac{N_\epsilon}{48 \pi^2 }\xi(\alpha^2-\beta)^2.
\end{eqnarray}
 Some limits of these renormalization relations can be easily checked. For example in the case of the MSM ($\alpha=\beta=\xi=1$) we see that none of these five couplings is strictly needed because of the renormalizability of the model. The case of the Higgsless EWChL corresponds to  $\alpha= \beta =0$ and consistently we find that   $ \gamma, \delta$ and 
$\eta$ do not need any renormalization and also we reproduce the well known results for the constants $a_4$ and $a_5$ \cite{Appelquist} in this case. Finally we have checked also that the renormalization  $a_4$ and $a_5$  agree with the corresponding ones found in \cite{Espriu:2013fia}.
 In terms of these renormalized couplings the elastic WBGB amplitude reads
\begin{eqnarray} \label{Arenorm}
A(s, t, u) & = & \frac{s}{v^2} (1 - \xi \alpha^2)+\frac{ 4}{v^4} [2 a^r_5(\mu) s^2 + a^r_4(\mu) (t^2 + u^2)] \\  \nonumber    
& + &\frac{1}{16 \pi^2 v^4}\left(\frac{1}{9} (14 \xi^2 \alpha^4 - 10 \xi \alpha^2 - 18 \xi^2 \alpha^2 \beta  + 9 \xi^2 \beta^2 + 
        5 ) s^2
     + \frac{13}{18} (\xi \alpha^2 - 1)^2 (t^2 + u^2) \right.  \\ \nonumber  
       &  - & \frac{1}{2}  (2 \xi^2 \alpha^4 - 2 \xi \alpha^2 - 
        2 \xi^2 \alpha^2 \beta  + \xi^2 \beta^2 + 
        1)   s^2 \log\frac{-s}{\mu^2} \\  \nonumber   
       &  +& \frac{1}{12} (1-\xi \alpha^2 )^2 (s^2 - 3 t^2 - 
        u^2) \log\frac{-t}{\mu^2}        \\ \nonumber  
       &  + & \left.
  \frac{1}{12}   (1-\xi \alpha^2 )^2 (s^2 - t^2 - 3 u^2) \log\frac{-u}{\mu^2}
    \right)\ .
\end{eqnarray}
The inelastic $\omega\omega  \rightarrow \varphi\varphi$ amplitude is correspondingly
\begin{eqnarray} \label{Mrenorm}
M(s,t,u)  & = & \frac{\alpha^2-\beta}{f^2}s  + \frac{2 \delta^r(\mu)}{v^2f^2}s^2+ \frac{\eta^r(\mu)}{v^2f^2}(t^2+u^2) \nonumber  \\
& + & 
\frac{(\alpha^2-\beta)}{576\pi^2v^2f^2}
\left\{\left[72 -  88 \xi \alpha^2+ 16 \xi \beta  + 36 (\xi \alpha^2-1)\log\frac{-s}{\mu^2}\right.\right. \nonumber \\
& + &
\left.\left. 3 \xi (\alpha^2-\beta)\left(\log\frac{-t}{\mu^2}+\log\frac{-u}{\mu^2}\right)\right]s^2 \right. \nonumber  \\
& + & \xi (\alpha^2-\beta)\left(26-9\log\frac{-t}{\mu^2}-3\log\frac{-u}{\mu^2}\right)t^2  \nonumber  \\
& + & \left. \xi (\alpha^2-\beta)\left(26-9\log\frac{-u}{\mu^2}-3\log\frac{-t}{\mu^2}\right)u^2 \right\}
\end{eqnarray}
and finally the $\varphi \varphi  \rightarrow \varphi \varphi$ amplitude may be written as
\begin{eqnarray} \label{Trenorm}
T(s,t,u) & = & \frac{2\gamma^r(\mu)}{f^4}(s^2+t^2+u^2) \\ \nonumber
 &+& 
\frac{3\xi^2(\alpha^2-\beta)^2 }{32\pi^2v^4}
\left[ 2(s^2+t^2+u^2)-s^2\log\frac{-s}{\mu^2}-t^2\log\frac{-t}{\mu^2}
-u^2\log\frac{-u}{\mu^2}\right]\ .
\end{eqnarray}
Apparently,  the amplitudes in Eqs.~(\ref{Arenorm}), (\ref{Mrenorm}) and (\ref{Trenorm}) depend on the arbitrary scale $\mu$ through the log terms. 
However they also depend on $\mu$ implicitly through the renormalized couplings $a_4\dots \eta$. 

However,  as there is no wave or mass renormalization, the amplitudes must be observable and therefore $\mu$-independent; then we may require that their total derivatives with respect to $\log \mu^2$ vanish. Integrating the resulting (very simple) differential equations, we find the renormalization-group evolution equations for the different couplings which turn out to be

\begin{eqnarray} \label{RGE}
a_4^r (\mu)& = & a_4^r(\mu_0)- \frac{1}{192 \pi^2}(1-\xi \alpha^2)^2    \log\frac{\mu^2}{\mu_0^2}   \nonumber \\
a_5^r(\mu) & = & a_5^r(\mu_0)- \frac{1}{768 \pi^2} (2+5 \xi^2\alpha^4-4 \xi \alpha^2-6\xi^2\alpha^2\beta+3\xi^2\beta^2) \log\frac{\mu^2}{\mu_0^2}\nonumber \\
\gamma^r(\mu) & = & \gamma^r(\mu_0)-\frac{3}{64\pi^2}(\alpha^2-\beta)^2  \log\frac{\mu^2}{\mu_0^2}  \nonumber  \\
\delta^r(\mu) & = & \delta^r(\mu_0) +\frac{1}{192 \pi^2}(\alpha^2-\beta)(7\xi\alpha^2-\xi\beta-6)  \log\frac{\mu^2}{\mu_0^2} \nonumber  \\
\eta^r(\mu)  & = & \eta(\mu_0)- \frac{1}{48 \pi^2}\xi(\alpha^2-\beta)^2 \log\frac{\mu^2}{\mu_0^2}\ .
\end{eqnarray}
These equations allow to reexpress the amplitudes at any second scale. 
They are diagonal, so that the various coefficients do not enter the evolution equation for any other ones, a feature that will not persist at higher orders in perturbation theory.
Since no resonance beyond the Standard Model is presently known, there is no particularly natural renormalization scale $\mu$, so we will arbitrarily employ $\mu=1$ TeV. All NLO numerical couplings quoted in section~\ref{sec:numeric} below are to be understood as taken at this scale.

\section{Partial wave behavior in electroweak ChPT}\label{sec:partialwaves}
The unitarity properties of the three scattering amplitudes are best exposed in terms of the isospin- and spin-projected partial waves. 
For elastic WBGB scattering there are three custodial-isospin $A_I$  amplitudes ($I=0,1,2$) analogous to those in pion-pion scattering in hadron physics, 
\begin{eqnarray}
A_0(s, t, u)  & = &  3 A(s, t, u) + A(t, s, u) + A(u, t, s)  \\ \nonumber
A_1(s, t, u)  & = & A(t, s, u) - A(u, t, s)     \\ \nonumber
A_2(s, t, u)  & = & A(t, s, u) + A(u, t, s)\ .
\end{eqnarray}
We can then project them over definite orbital angular momentum (the WBGBs carry zero spin), and choose the normalization as
\begin{equation} \label{Jprojection}
A_{IJ}(s)=\frac{1}{64\,\pi}\int_{-1}^1\,d(\cos\theta)\,P_J(\cos\theta)\,
A_I(s,t,u)\ .
\end{equation}

These partial waves also accept a chiral expansion
\be
A_{IJ}(s)=A^{(0)}_{IJ}(s)+A^{(1)}_{IJ}(s)+... ,
\ee
where
\begin{eqnarray}\label{expandpartialwave}
\nonumber   A^{(0)}_{IJ}(s)     & = & K s   \\
    A^{(1)}_{IJ}(s) & = &  s^2\left( B(\mu)+D\log\frac{s}{\mu^2}+E\log\frac{-s}{\mu^2}\right) \ .
\end{eqnarray}
The constants $K$, $D$ and $E$ and the function $B(\mu)$ depend on the different channels $IJ=00,11,20,02$ as shown below 
and we will use the same notation for the inelastic and pure-$\varphi$ scattering reactions.

As $A_{IJ}(s)$ must be scale independent  we have
\begin{equation}
B(\mu)=
 B(\mu_0)+(D+E)\log\frac{\mu^2}{\mu_0^2}\ ;
 \end{equation}
For elastic $\omega\omega$ scattering, 
this $B$ function is linear in the NLO chiral constants (with certain proportionality coefficients $p_4$ and $p_5$ that can be read off Eq.~(\ref{partial00}) and following)
\be 
\label{Bfunction}
B(\mu)=B_0 +p_4 a_4(\mu)+p_5 a_5(\mu)\ ,
\ee
where from now on we omit the superindices $r$ on the renormalized coupling constants for simplicity.

A direct evaluation of the integral in Eq.~(\ref{Jprojection}) substituting the renormalized amplitude obtained in Eq.~(\ref{Arenorm}) for the $\omega\omega \rightarrow \omega\omega$ process produces the following auxiliary $K$, $D$, $E$ constants and $B(\mu)$ functions. 

 For the scalar-isoscalar channel with $IJ=00$,
\begin{eqnarray}{}\label{partial00}
\nonumber   K_{00} & = & \frac{1}{16 \pi v^2} (1-\xi \alpha^2) \\
 \nonumber   B_{00}(\mu) & = &\frac{ 1}{9216 \pi^3 v^4} [101 + 
   768 (7 a_4(\mu) + 11 a_5(\mu)) \pi^2 + \xi (169 \alpha^4 \xi + 
      68 \beta^2 \xi - 2 \alpha^2 (101 + 68 \beta \xi))]\\
 \nonumber   D_{00} & = & -\frac{1}{4608\pi^3v^4} [7 + \xi (10 \alpha^4 \xi + 3 \beta^2 \xi - 
      2 \alpha^2 (7 + 3 \beta \xi))]\\
 E_{00} & = & -\frac{1}{256\pi^3v^4} [(1-\xi \alpha^2)^2  +\frac{3}{4} \xi^2 (\alpha^2-\beta) ^2    )]\ .
\end{eqnarray}
For the vector isovector $IJ=11$ amplitude,
\begin{eqnarray}{}\label{partial11}
\nonumber   K_{11} & = & \frac{1}{96 \pi v^2} (1-\xi \alpha^2) \\
 \nonumber   B_{11}(\mu) & = & \frac{1}{110592\pi^3 v^4}[8 + 
   4608 (a_4(\mu) - 2 a_5(\mu)) \pi^2 - \xi (67 \alpha^4 \xi + 
      75 \beta^2 \xi + 2 \alpha^2 (8 - 75 \beta \xi))]\\
 \nonumber   D_{11} & = & \frac{1}{9216\pi^3v^4}      [1 + \xi (4 \alpha^4 \xi + 3 \beta^2 \xi - 
     2 \alpha^2 (1 + 3 \beta \xi))] \\
  E_{11} & = & -\frac{1}{9216\pi^3v^4}  (1-\xi \alpha^2)^2\ .
\end{eqnarray}
For the scalar isotensor $IJ=20$:
\begin{eqnarray}{}\label{partial20}
\nonumber   K_{20} & = & -\frac{1}{32 \pi v^2} (1-\xi \alpha^2) \\
 \nonumber   B_{20}(\mu) & = & \frac{1}{18432 \pi^3 v^4}
 [91 + 3072 (2 a_4(\mu) + a_5(\mu)) \pi^2 + 
  7 \xi (17 \alpha^4 \xi+ 4 \beta^2 \xi - 
     2 \alpha^2 (13 + 4 \beta \xi))]\\
 \nonumber   D_{20} & = & -\frac{1}{9216\pi^3v^4} [11 + \xi (17 \alpha^4 \xi+ 6 \beta^2 \xi - 
     2 \alpha^2 (11 + 6 \beta \xi))]\\
  E_{20} & = & -\frac{1}{1024\pi^3v^4}(1-\xi \alpha^2)^2  \ 
\end{eqnarray}
and finally for the tensor isoscalar $IJ=02$,
\begin{eqnarray}{}\label{partial02}
\nonumber   K_{02} & = & 0 \\
 \nonumber   B_{02}(\mu) & = & \frac{1}{921600 \pi ^3 v^4}
 [320 + 15360 (2 a_4(\mu) + a_5(\mu)) \pi^2 + \xi (397 \alpha^4 \xi + 
     77 \beta^2 \xi - 2 \alpha^2 (320 + 77 \beta \xi))] \\
 \nonumber   D_{02} & = & -\frac{1}{46080\pi^3v^4}[10 + \xi (13 \alpha^4 \xi + 3 \beta^2 \xi- 
     2 \alpha^2 (10 + 3 \beta \xi))] \\
  E_{02} & = & 0\ .
\end{eqnarray}

Since the ``Higgs'' boson has zero custodial isospin, 
 the $\omega\omega  \rightarrow \varphi\varphi$ and $\varphi \varphi  \rightarrow \varphi \varphi$ reactions only proceed in the isospin zero channel  $I=0$. In the first, inelastic, case we have $M_0(\omega\omega  \rightarrow \varphi\varphi)=\sqrt{3} M(s,t,u)$ and  for the scalar-scalar interaction, $T_0(\varphi \varphi  \rightarrow \varphi \varphi)= T(s,t,u)$. 
The chiral expansions equivalent to the $\omega\omega$ elastic one in Eq.~(\ref{expandpartialwave})
are now
\begin{eqnarray}
\nonumber   M_{J}(s)     & = & K' s+  s^2\left( B'(\mu)+D'\log\frac{s}{\mu^2}+E'\log\frac{-s}{\mu^2}\right) \dots \\
    T_{J}(s)     & = & K'' s  +  s^2\left( B''(\mu)+D''\log\frac{s}{\mu^2}+E''\log\frac{-s}{\mu^2}\right) \dots
\end{eqnarray}
(with $J$ subindex omitted in the constants).  The functions $B'(\mu)$ and $B''(\mu)$ are in all analogous to
$B(\mu)$ as defined in Eq.(\ref{Bfunction}), but with the constants $a_4$, $a_5$ renormalizing the elastic $\omega\omega$ channel being substituted by $\delta$, $\eta$ (for $B'$) and $\gamma$ (for $B''$) involving the $\varphi$ boson.

In consequence we find for the $\omega\omega  \rightarrow \varphi\varphi$ 
partial waves $M_{J}$, starting by the scalar one,
\begin{eqnarray}{} \label{Mscalar}
\nonumber   K'_{0} & = & \frac{\sqrt{3}}{32 \pi f^2} (\alpha^2-\beta) \\
 \nonumber   B'_{0}(\mu) & = & \frac{\sqrt{3}}{16\pi v^2f^2} \left(\delta(\mu)+\frac{\eta(\mu)}{3}\right)   -  \frac{\sqrt{3}(\alpha^2-\beta)}{18432\pi^3 v^2f^2}(71\xi\alpha^2+\xi\beta-72)    \\
 \nonumber   D'_{0} & = & - \frac{\sqrt{3}(\alpha^2-\beta)^2}{9216\pi^3 f^4}   \\  
  E'_{0} & = &  -\frac{\sqrt{3}(\alpha^2-\beta)}{512\pi^3 v^2f^2}(1-\xi\alpha^2)                                   
\end{eqnarray}
while for the tensor $M_{2}$ channel
\begin{eqnarray}{} \label{Mtensor}
\nonumber   K'_{2} & = & 0 \\
 \nonumber   B'_{2}(\mu)& = & \frac{\eta(\mu)}{160\sqrt{3}\pi v^2f^2}  +\frac{83(\alpha^2-\beta)^2}{307200\sqrt{3}\pi^3 f^4}    \\
 \nonumber   D'_{2} & = & - \frac{(\alpha^2-\beta)^2}{7680\sqrt{3}\pi^3 f^4} \\ 
  E'_{2} & = &  0\  .                                
\end{eqnarray}
Finally for the  $\varphi \varphi \rightarrow  \varphi\varphi$ reaction the $T_{0}(s)$ scalar partial-wave amplitude is given by the set of constants

\begin{eqnarray}{}\label{Tscalar}
\nonumber   K''_{0} & = & 0 \\
 \nonumber   B''_{0}(\mu) & = & \frac{10\gamma(\mu) }{96\pi f^4}   +  \frac{(\alpha^2-\beta)^2}{96\pi^3 f^4}   \\
 \nonumber   D''_{0} & = & - \frac{(\alpha^2-\beta)^2}{512\pi^3 f^4}  \\ 
  E''_{0} & = &  -   \frac{3(\alpha^2-\beta)^2}{1024\pi^3 f^4}   
\end{eqnarray}{}  
and the tensor $T_{2}$ in turn by
\begin{eqnarray} \label{Ttensor}
\nonumber   K''_{2} & = & 0 \\
 \nonumber   B''_{2}(\mu) & = & \frac{\gamma(\mu) }{240\pi f^4}   +  \frac{77(\alpha^2-\beta)^2}{307200\pi^3 f^4}   \\
 \nonumber   D''_{2} & = &  -\frac{(\alpha^2-\beta)^2}{5120\pi^3 f^4}  \\ 
  E''_{2} & = &    0 \ .
  \end{eqnarray}         
The $\mu$-invariance of all the above partial waves is easy to check by
substituting the $\mu$-evolution of the renormalized couplings in Eq.~(\ref{RGE}).

The partial-wave amplitudes $A_{IJ}(s)$, $M_{J}(s)$ and $T_{J}(s)$ are all analytical functions of complex Mandelstam-$s$, having the proper left and right (or unitarity) cuts, shortened to LC and RC respectively.
The physical values of their argument are $s=\it{E}_{\rm CM}^2+i\epsilon$ (i.e. on the upper lip of the RC), where $\it{E}_{\rm CM}$ is the total energy in the center of mass frame.
For these physical $s$ values,  exact unitarity requires a set of non-trivial relations between the different partial waves that we now spell out.

 For $I=0$ and either of $J=0$, $J=2$, where channel coupling is possible,
\begin{eqnarray}{}
 \Imag A_{0J} &  = & \lvert A_{0J} \rvert ^2 +  \lvert M_J \rvert ^2 \\
 \nonumber    \Imag M_J &  = & A_{0J} M_{J}^*+  M_J T_J^* \\
 \nonumber    \Imag T_J &  = & \lvert M_J \rvert ^2+  \lvert T_J \rvert ^2\   .
\end{eqnarray}         
These relations are not exactly respected by perturbation theory, but are instead satisfied only to one less order in the expansion than kept in constructing the amplitude. At the one-loop level one has
\begin{eqnarray}{}
\nonumber \Imag A^{(1)}_{0J} &  = &\lvert A^{(0)}_{0J}\rvert^2+  \lvert M^{(0)}_{J}\rvert ^2 \\
 \nonumber   \Imag M^{(1)}_{J} &  = & A^{(0)}_{0J} M^{(0)}_{J}+ 
 M^{(0)}_{J}T^{(0)}_{J} \\
 \nonumber    \Imag T^{(1)}_{J} &  = & \lvert M^{(0)}_J\rvert ^2 +  \lvert T^{(0)}_{J}\rvert^2   .
\end{eqnarray}    
     
For the remaining channels with $I=J=1$ and $I=2$, $J=0$ the $\omega\omega \rightarrow \omega\omega$ reaction is elastic and the unitarity condition is just
\be \label{unitarity1channel}
 \Imag A_{I J}   =  \lvert A_{IJ} \rvert ^2  \ \ \ I\neq 0
\ee
and at the NLO perturbative level,
\be
 \Imag A^{(1)}_{I J}   =  \lvert A^{(0)}_{IJ} \rvert ^2  \ \ \ I\neq 0\ .
\ee

There are in all eight independent one-loop perturbative relations,
that can also be obtained by applying the Landau-Cutkosky cutting rules 
and directly checked in each of  the partial waves for the three reactions, providing  a very good, non-trivial check of our amplitudes.

\section{Phenomenology of the partial waves}\label{sec:numeric}
\subsection{Numerical evaluation}

In this section we evaluate all partial waves with the constants in Eq.~(\ref{partial00}) and following and expose their dependence on the LO parameters that separate them from the Standard Model, and on the NLO parameters as well.
Generically, the partial waves (whether we plot the real, the imaginary part or the modulus) will correspond to the $OY$ axis and be denoted by $t$, while the $OX$ axis is the squared physical cm energy $s=E_{\rm cm}^2$.

\begin{figure}[h]
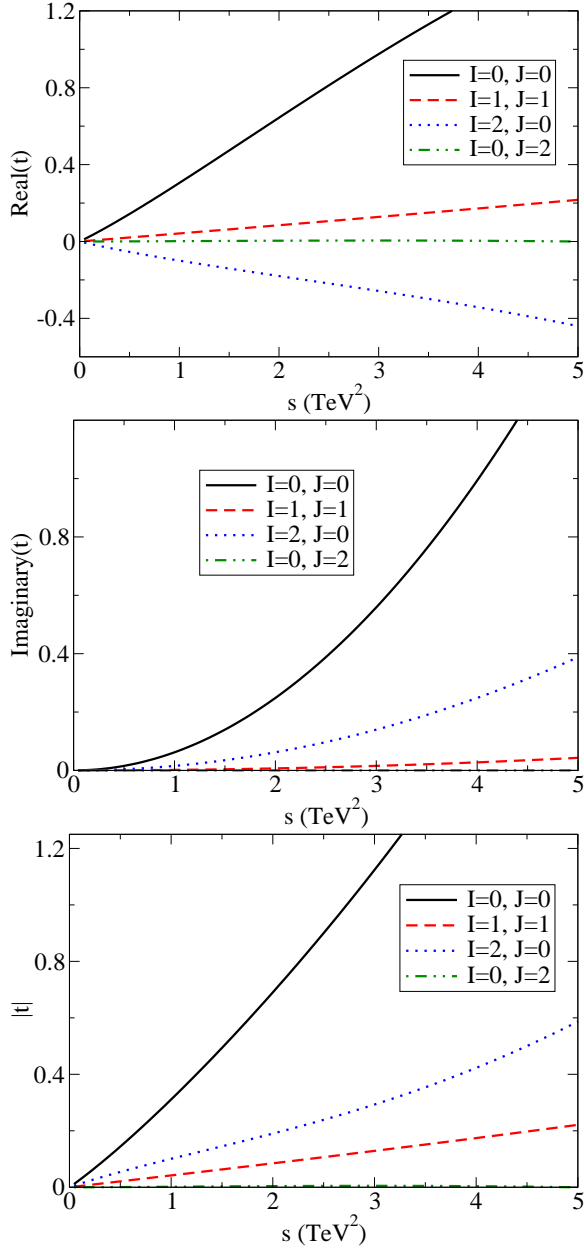

\begin{center}
\includegraphics*[width=.5\textwidth]{FIGS.DIR/Realpertnoaes.eps}\\
\includegraphics*[width=.5\textwidth]{FIGS.DIR/Imagpertnoaes.eps}\\
\includegraphics*[width=.5\textwidth]{FIGS.DIR/Modpertnoaes.eps}
\end{center}
\caption{\label{fig:wwnoaes} From top to bottom: real part, imaginary part, and modulus of the elastic $\omega\omega\to \omega\omega$ scattering amplitude to one loop. Here $f=500$ GeV (approximately 2$v$), and the NLO constants are chosen to be $a_4=a_5=0$
at a scale $\mu=1$ TeV. We show the four NLO non-vanishing partial waves $A_{IJ}$.}
\end{figure}

In figure~(\ref{fig:wwnoaes}) we have plotted the elastic $\omega\omega\to
\omega\omega$ amplitude without NLO constants, by setting $a_4=a_5=0$ at $\mu=1$ TeV (scale also chosen in all examples to follow). Also $f$ has been chosen at 500 GeV, which is about $2v$, and to avoid channel coupling we have kept $\alpha^2=\beta=1$ (Standard Model values) so that all the $(\alpha^2 - \beta)$ factors vanish.

First we observe the real part of the amplitude (top plot in the figure). We conclude that just like in low-energy hadron physics, the $IJ=00$ wave is strongly attractive, the $IJ=11$ (without NLO constants) mildly attractive, the $02$ wave negligible in the low-energy region, and the $IJ=20$ wave is actually repulsive.

Next we turn to the imaginary part (middle plot) and modulus (bottom plot). 
It is plain that unitarity is badly violated at a scale between 2 and 3 TeV (for this modest value of $f$) because the modulus of $A_{00}$ exceeds 1, which is not possible according to Eq.~(\ref{unitarity1channel}). But moreover, the equation is not well satisfied even for much smaller scales. This is a handicap of perturbation theory.

\begin{figure}[h]
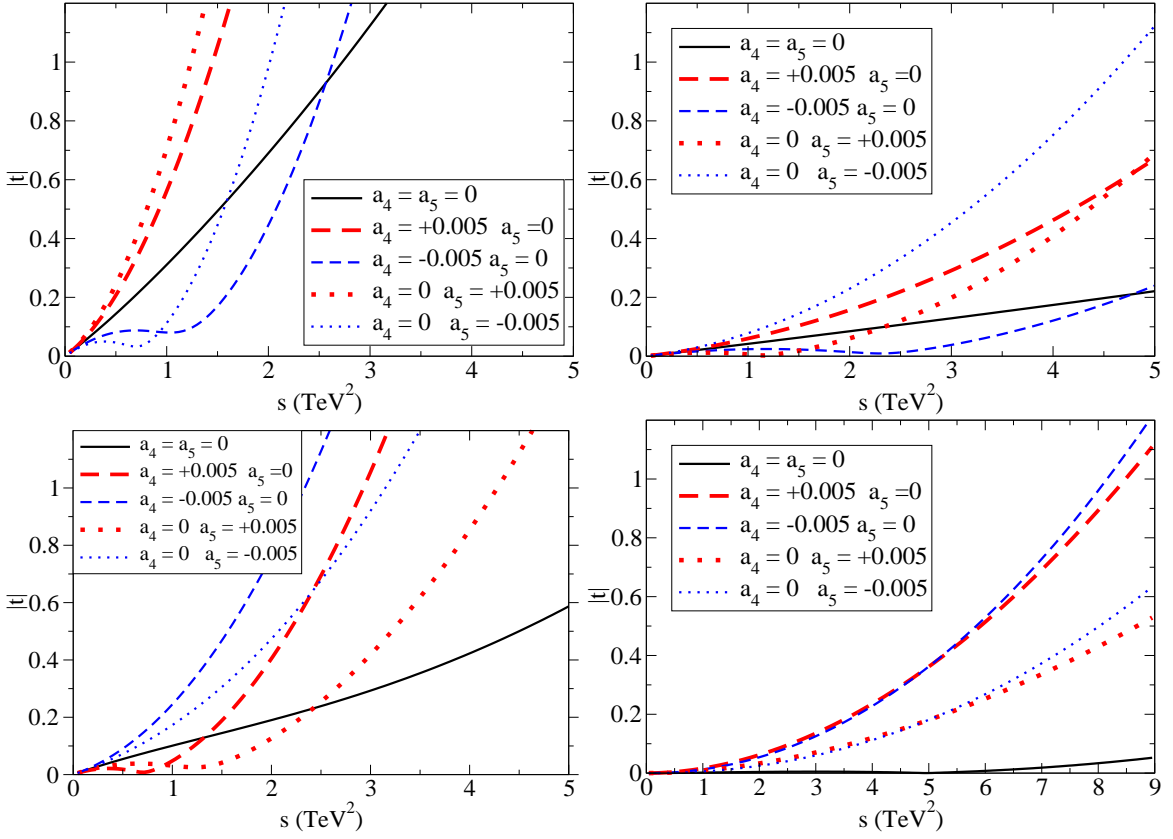

\includegraphics*[width=.5\textwidth]{FIGS.DIR/Modpert00.eps}
\includegraphics*[width=.5\textwidth]{FIGS.DIR/Modpert11.eps}\\
\includegraphics*[width=.5\textwidth]{FIGS.DIR/Modpert20.eps}
\includegraphics*[width=.5\textwidth]{FIGS.DIR/Modpert02.eps}
\caption{\label{wwaes} For $f=500$ GeV, we take a non-zero (positive or negative) $a_4$ or $a_5$ at $\mu=1$ TeV and plot the modulus of the partial wave amplitudes for elastic $ww\to ww$ scattering. In clockwise sense from the top left, we show
$\ar A\ar_{00}$, $\ar A\ar_{11}$, $\ar A\ar_{02}$, $\ar A\ar_{20}$.}
\end{figure}

We now switch-on $a_4$ and $a_5$ within the range of
values explored in reference~\cite{Espriu:2012ih}, and plot the results in figure~\ref{wwaes}. 

In agreement with that reference, we find that positive values of $a_4$ or $a_5$ enhance the $IJ=00$ channel at low energy, while negative values suppress it. The $11$ vector amplitude is enhanced by positive $a_4$ and negative $a_5$, while the $20$ isotensor one is larger for negative $a_4$ or $a_5$. The $IJ=02$ amplitude seems too small in the low-energy region to be of much use in early experiments for small $a_4$ and $a_5$, but it is sensitive to the NLO terms.

\begin{figure}[h]
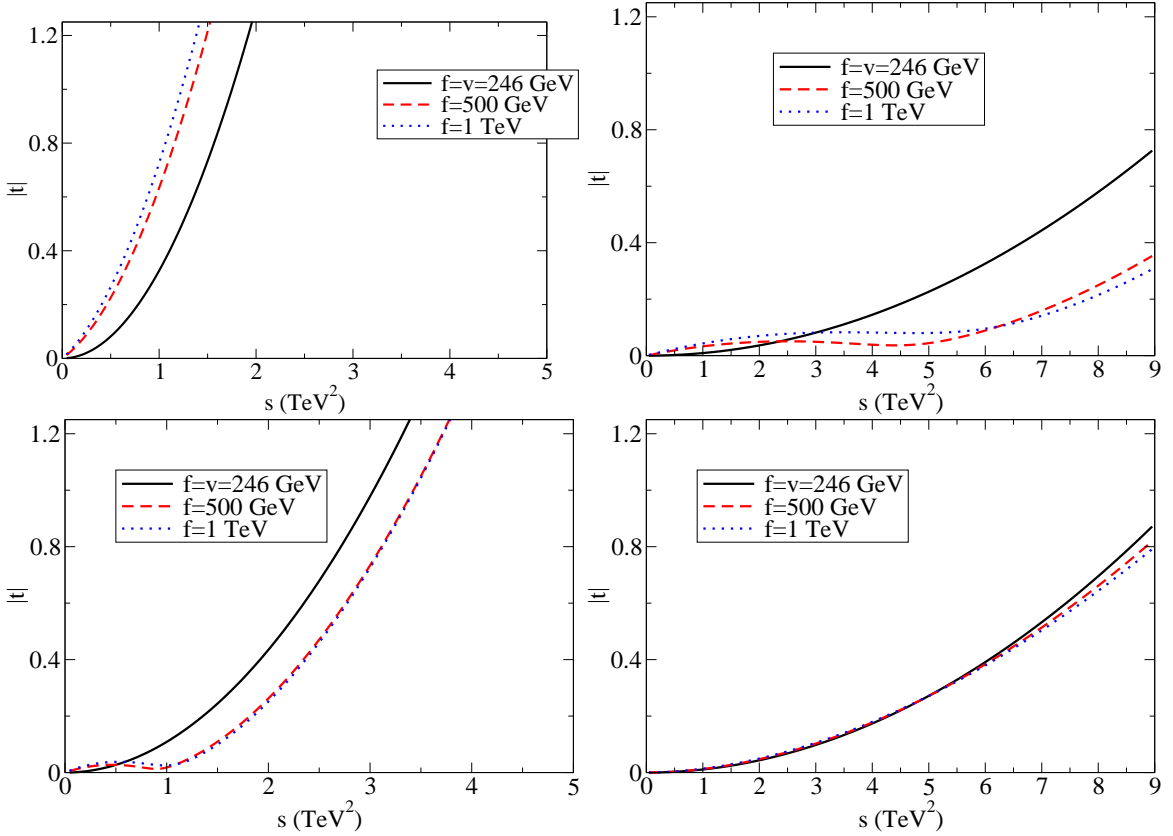

\includegraphics*[width=.5\textwidth]{FIGS.DIR/Mod00Varyf.eps}
\includegraphics*[width=.5\textwidth]{FIGS.DIR/Mod11Varyf.eps}\\
\includegraphics*[width=.5\textwidth]{FIGS.DIR/Mod20Varyf.eps}
\includegraphics*[width=.5\textwidth]{FIGS.DIR/Mod02Varyf.eps}
\caption{\label{wwvaryf} For fixed $a_4=a_5=0.0025$ at $\mu=1$ TeV, we vary $f$ as indicated and plot the modulus of the perturbative partial wave amplitudes for elastic $ww\to ww$ scattering. In clockwise sense from the top left, we show
$\ar A_{00}\ar$, $\ar A_{11}\ar$, $\ar A_{02}\ar$, $\ar A_{20}\ar$.}
\end{figure}

We now fix $a_4$ and $a_5$ to 0.0025 at the same scale of 1 TeV and note the variation of the amplitudes respect to $f$ in figure~\ref{wwvaryf}.

\newpage

$\lvert A_{00}\rvert$, in the top left plot, is seen to shoot more rapidly for larger $f$, implying the generic $\omega\omega$ interaction will be stronger yet. The effect is opposite for the tensor amplitude in the bottom right panel, $\lvert A_{02}\rvert$, presenting a smaller modulus. The other two amplitudes are initially larger at smaller energies, but as $s$ increases the trend changes and they become less prominent for larger $f$.

\begin{figure}[h]
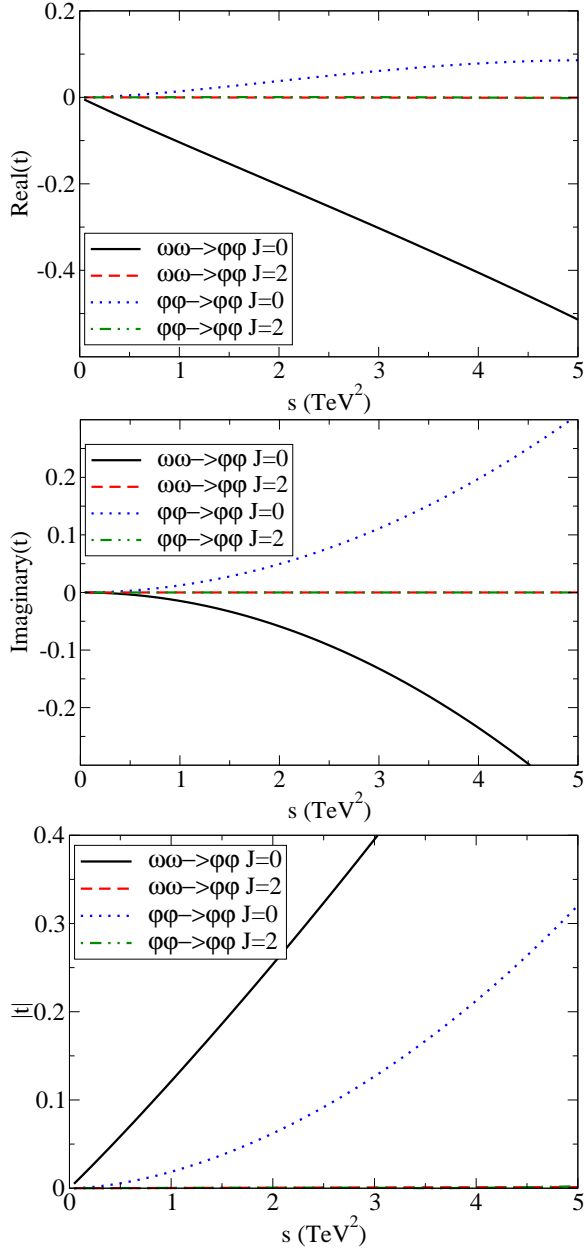

\begin{center}
\includegraphics*[width=.5\textwidth]{FIGS.DIR/Realpertnogamas.eps}\\
\includegraphics*[width=.5\textwidth]{FIGS.DIR/Imagpertnogamas.eps}\\
\includegraphics*[width=.5\textwidth]{FIGS.DIR/Modpertnogamas.eps}
\end{center}
\caption{\label{fig:phiphinogamas} From top to bottom: real part, imaginary part, and modulus of the elastic $\varphi\varphi\to \varphi\varphi$ and cross-channel $\omega\omega\to \varphi\varphi$ scattering amplitude to one loop. Here $f=350$ GeV (somewhat larger than $v$), $\alpha=1.1$, $\beta=2$, and the NLO constants are chosen as $\gamma\ , \delta \ , \eta(\mu=1{\rm TeV})=0$.}
\end{figure}

Figure~\ref{fig:phiphinogamas} shows the inelastic $\omega\omega\to\varphi\varphi$ and elastic $\varphi\varphi\to \varphi\varphi$ for parameters $f=350$ GeV, $\alpha=1.1$ (perfectly allowed by current LHC bounds~\cite{bounds}), $\beta=2$ (unconstrained at the LHC) and all three NLO parameters $\gamma$, $\delta$, $\eta$ set to zero at 1 TeV. Because $\alpha^2-\beta$ is negative with this parameter choice, the real part of $M_{0}$ is also negative, while  $Re T_0$ remains positive due to the factor appearing squared. More than in elastic $\omega\omega$ scattering, the $J=2$ amplitudes are completely negligible.

\begin{figure}[h]
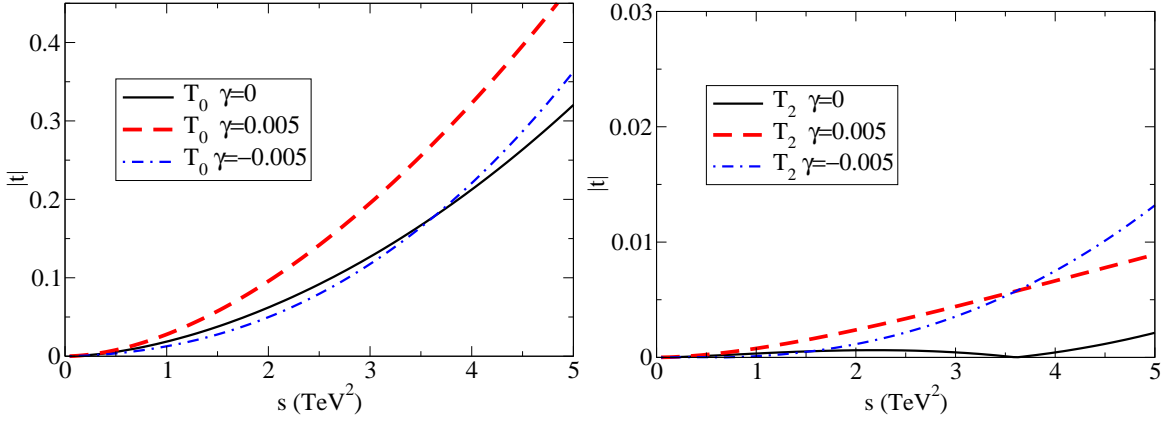

\includegraphics*[width=.5\textwidth]{FIGS.DIR/Modpert_T0.eps}
\includegraphics*[width=.5\textwidth]{FIGS.DIR/Modpert_T2.eps}
\caption{\label{fig:phiphigamas} $\varphi\varphi$ elastic scattering in the presence of the NLO $\gamma$ parameter with $\mu=1$ TeV. 
Left: modulus of the scalar partial-wave. Right: modulus of the tensor partial-wave. Note the very different scale.
}
\end{figure}

This is also the case for the $\varphi\varphi\to \varphi\varphi$ tensor amplitude in figure~\ref{fig:phiphigamas}  showing the sensitivity to including the $\gamma$ parameter with a small value of $\pm 0.005$; obviously if a tensor resonance exists that couples to this $T_{0}$ channel, it will entail a large value of $\gamma$. The scalar amplitude is more commensurate with others, yet keeping in mind that it is very dependent on $(\alpha^2-\beta)^2$. The effect of a positive $\gamma$ is to enhance (negative $\gamma$, to decrease) the amplitude at very low scales.

\begin{figure}[h]
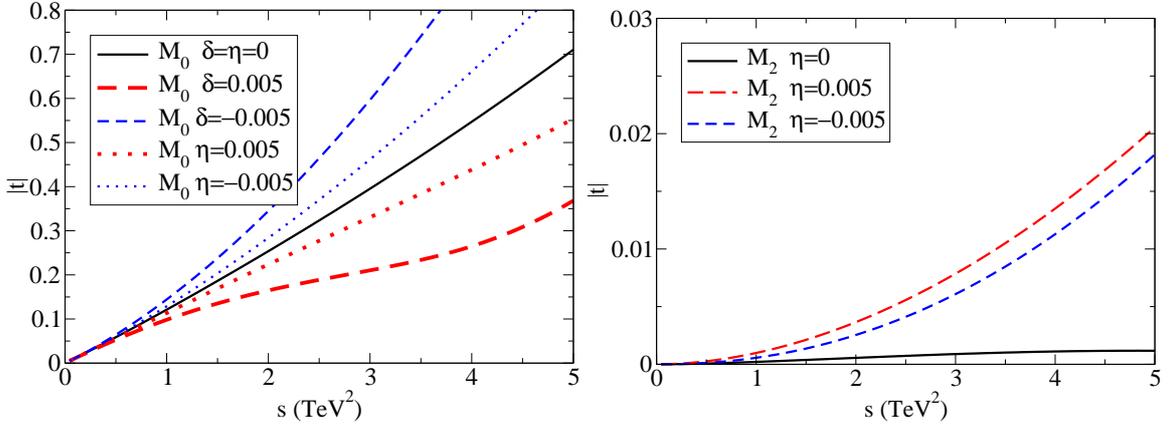

\includegraphics*[width=.5\textwidth]{FIGS.DIR/ModpertM0.eps}
\includegraphics*[width=.5\textwidth]{FIGS.DIR/ModpertM2.eps}
\caption{\label{fig:coupleddeltaeta} $\omega\omega\to \varphi\varphi$ 
channel-coupling amplitude in the presence of the NLO $\delta$ and $\eta$ parameters taken at $\mu=1$ TeV, alternatively.
Left: modulus of the scalar partial-wave. Right: modulus of the tensor partial-wave. Note the very different scale.
}
\end{figure}

In figure~\ref{fig:coupleddeltaeta} we plot the moduli of the $\omega\omega\to \varphi\varphi$ scalar and tensor amplitude showing the effect of adding either $\delta=\pm 0.005$ or $\eta=\pm 0.005$. 
The tensor piece $M_2$ is only affected by $\eta$ as per Eq.~(\ref{Mtensor}). The scalar amplitude on the left panel is on the other hand influenced by both, and becomes larger for either of $\delta$ or $\eta$ taking a negative value.
\newpage

\newpage
\subsection{Experimental extraction of parameters}
In view of these results, an experimental programme to measure the parameters of the EWSBS and check them against the Minimal Standard Model from ``low'' energy data in the TeV region or below would  start by a partial-wave analysis of $W_LW_L$ spectra, where one would hope to be able to fit 
$a_4$, $a_5$, $f$ and $\beta$ (setting $\alpha=1$).  The tensor wave being very small a priori, one would resort to the scalar-isoscalar, vector-isovector, and scalar-isotensor final states, selected by the charge combinations of the $W$'s. If $A_{02}$ is nevertheless found to be large, this would immediately point out to important NLO contact terms. 

To proceed, one would first attempt an extraction of the leading order ($\propto s$) scalar amplitude (see Eq.~(\ref{partial00}) ) whose slope gives access to $f$. Then the three NLO ($\propto s^2$) partial waves in Eqs.(\ref{partial00}), (\ref{partial11}), (\ref{partial20}) give access to different linear combinations of $a_4$, $a_5$ and $\beta$, so obtaining the slopes of the $s^2$ terms in the spectra allows their isolation.

In the absence of a $\varphi\varphi$ spectrum, a unitarity analysis of the partial $W_LW_L$ waves may reveal the leak of probability to that unmeasured channel. The slope of the LO term ($\propto s$) is an independent measure of $\beta/f^2$.  The  slope of the NLO $s^2$ term in Eq.~(\ref{Mscalar}) then gives access to the combination $\delta + \eta/3$. Separately measuring $\eta$ (and thus $\delta$) seems quite hopeless because it requires to separate the tiny tensor $M_2$ channel. Unless the BSM spectrum contains a tensor resonance in the TeV scale, this will be heroic.

One can do little else unless a $\varphi\varphi$ two-scalar boson spectrum becomes available. In such case,
one may access the $\gamma$ parameter directly from the scalar amplitude at NLO, while fitting simultaneously $\delta$ and $\eta$.

\begin{figure}[h]
\begin{center}
\includegraphics*[width=8.7cm]{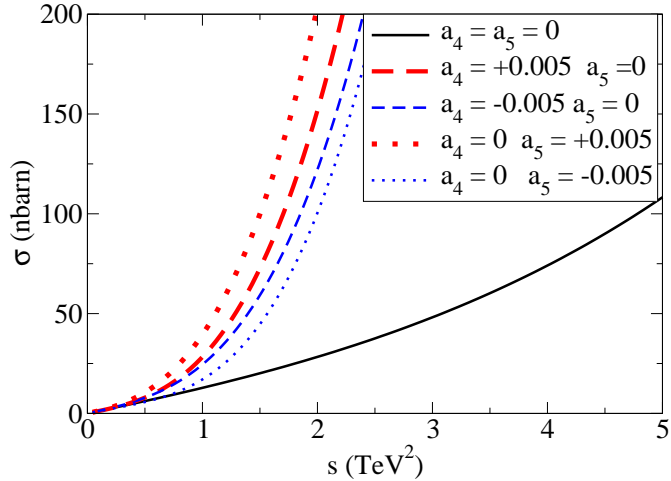}
\end{center}
\caption{\label{fig:cross} Cross-section for $\omega\omega \to \omega\omega$  resulting from evaluating Eq.~(\ref{crosssection}) with the amplitudes in
figure~\ref{fig:cross} and the parameters there described (taken again at $\mu=1$ TeV).}
\end{figure}

It remains to comment that, although we have been speaking of ``strong'' interactions in case the parameters of the low-energy Lagrangian density separate from the MSM, the cross-sections are rather small because of the $1/s$ flux factor. For example, the $\omega\omega\to\omega\omega$  cross-section can be expressed as
\be\label{crosssection}
\sigma(s) = \frac{64\pi}{s} \sum_{IJ} (2I+1)(2J+1) \lvert A_{IJ}\rvert^2 \ ;
\ee
the $1/s$ factor sets the scale at 1 TeV$^{-2}\simeq 0.39$ nbarn (increased to a meager 78 nbarn with the $64\pi$ factor).
For example, we can use the amplitudes from figure~\ref{wwaes} with the same parameters there indicated to plot the elastic $\omega\omega \to \omega\omega$ cross section in figure~\ref{fig:cross}, that shoots up rapidly for essentially any $a_4$ or $a_5$ enhancing BSM physics, but remains relatively small when compared with hadronic cross sections (of order 70 mbarn at the LHC, five orders of magnitude larger).

What ``strong interactions'' means in this context is that the $A_{IJ}$ amplitudes have moduli of order 1.
Then, for example, Watson's final state theorem applies due to rescattering, and the phases of the $W_LW_L$ or $\varphi\varphi$ production amplitudes should be the same as the phases of the elastic amplitudes (that are not directly accessible since we do not have asymptotic beams of  these unstable particles).

\subsection{Tree-level Higgs scattering: terms proportional to $m_\varphi^2$ }\label{subsec:tree}
We have been working in the chiral limit with $m_\varphi^2 \simeq m_W^2 \simeq m_Z^2 \simeq 0$. This is
a theoretical limit that may bear resemblance with reality in the energy region $s\simeq 1$ TeV$^2$ where squared momenta are significantly larger than masses, but it is perhaps useful to briefly assess the size of the terms
neglected.

We focuse on $\varphi \varphi \to \varphi \varphi$ elastic scattering, because our chiral amplitude vanishes at $O(s)$ with the series starting at $O(s^2)$ (see Eq.~(\ref{Tscalar}) where $K=0$) so one expects maximum sensitivity to the correction.

Our amplitude, at a simple reference point such as $\mu^2=s=1$ TeV$^2$ can be written as 
\ba \label{ourstocompare}
 T_0 \left(s=1\ {\rm TeV^2}\right) &=& \frac{1\ {\rm TeV}^4}{96\pi v^4} \xi^2\left(10\gamma(1{\rm TeV}) +\frac{\alpha^2-\beta}{\pi^2}\right) \\ \nonumber
 &=& 0.905\xi^2 \left(10\gamma +\frac{\alpha^2-\beta}{\pi^2}\right)
\ea

In the first place, if we consider instead the Higgs self-coupling potential in the Standard Model,
\be
V^{self} = \frac{m_\varphi^2}{2v}\varphi^2  + \frac{m_\varphi^2}{8v^2} \varphi^4\ ,
\ee
in the absence of new physics, the amplitude (heretofore vanishing due to our taking the chiral limit) would be, after projecting over $J=0$,
\ba
T^{self}_0 = - \frac{3m_\varphi^2}{32\pi v^2}\left(
1 + \frac{3m_\varphi^2}{s-m_\varphi^2} + 2\times \frac{3m_\varphi^2}{s-4m_\varphi^2}\log\left(\frac{m_\varphi^2}{s-3m_\varphi^2}\right) \right) \ .
\ea
At $s=1$ TeV$^2$, this is numerically equal to $-7.7(1+0.048+2\times 0.205)=-1.12\ 10^{-2}$.

Comparing with Eq.~(\ref{ourstocompare}), we see that the Standard Model Higgs self-couplings are negligible respect to the BSM ones at a scale of 1 TeV when
\be
\xi^2 (\alpha^2-\beta) \gg 0.12 \ .
\ee
As discussed in~\cite{Delgado:2013loa}, this is phenomenologically viable (essentially, $\beta$ is unconstrained to date). 

Even when $\alpha^2=\beta=1$, the other term in eq.~\ref{ourstocompare} can also dominate the scattering if
\be
10\pi^2\xi^2\gamma \gg 0.12\ ,
\ee
as Standard Model Higgs self-couplings would then be negligible. In this case $\alpha^2=\beta=1$  (see sec.~\ref{sec:partialwaves}), so the channels decouple; still,  the interactions are strong.

Nevertheless, if the separation from the SM is small so that $\alpha^2 \simeq \beta$, $\xi\simeq 1$ and $\gamma\lesssim 1.2\cdot 10^{-3}$, a
phenomenological analysis should keep the SM couplings (this is akin to keeping the pion-mass terms in chiral perturbation theory for low-energy QCD, and routinely done).

As we proceed to consider operators beyond the standard model in the Higgs sector, we encounter two more of dimension six~\cite{Jenkins:2013zja}, that involve two derivatives of the Higgs doublet field,
\ba
Q_{H\Box} &:=& \left(H^\dagger H\right)\Box\left(H^\dagger H\right)
\\ \nonumber
Q_{HD} &:=&  \left(H^\dagger D_\mu H\right)^* \left(H^\dagger D^\mu H\right)\ .
\ea
After substituting the real physical field $\varphi$ and neglecting the coupling to the transverse gauge bosons, 
they reduce to
\ba
Q_{\varphi\Box} &=& \varphi^2 \Box \varphi^2
\\ \nonumber
Q_{\varphi \partial} &=&  \left(\varphi \partial_\mu \varphi\right) \left(\varphi \partial^\mu \varphi\right)\ .
\ea
These last two operators are related by use of Green's first identity,
\be
Q_{\varphi \Box} = -4 Q_{\varphi \partial} + \ {\rm boundary\ term}\ .
\ee
In this paragraph we explore the addition of a term proportional to $Q_{\varphi \partial}$ to the effective Lagrangian density in Eq.~(\ref{genericLagrangian}).

One can discuss the normalization of the operator, whether $\frac{\gamma_6}{4 f^2} Q_{\varphi \partial}$ 
or $\frac{v^2 \gamma_6}{4 f^4} Q_{\varphi \partial}$ should be taken in the effective Lagrangian, with a certain coefficient $\gamma_6$. Irrespective of this, the scattering amplitude $\varphi_1\varphi_2\to \varphi_3\varphi_4$ can be easily calculated, yielding
\be
iT_{Q_{\varphi\partial}} = -4i \frac{\gamma_6}{4 f^2} \left(\xi^2\right) 
\left(p_1p_2-p_1p_3-p_1p_4-p_2p_3-p_2p_4+p_3p_4\right) 
\ee
where the prefactor of 4 is combinatoric (hence the 1/4 in the normalization) and $\xi^2$ depends on how one decides to normalize the operator. Eliminating the momenta in terms of the Mandelstam variables,
\be
iT_{Q_{\varphi\partial}} = -i \frac{\gamma_6}{f^2} \left(\xi^2\right) 
\left(s+t+u-6m_\varphi^2 \right) 
\ee
we see that 
\be
T_{Q_{\varphi\partial}} = \gamma_6 \frac{2m_\varphi^2}{f^2} \left(\xi^2\right) \ .
\ee

Nominally, the operator is zero in the chiral limit $m_\varphi\to 0$ and thus negligible in the TeV region 
unless the unknown coefficient $\gamma_6$ is not of natural size (in the Standard Model, of course, $\gamma_6=0$).

In conclusion, because the operator $Q_{\varphi\partial}$ is of one more order (at least) in the $1/f$ counting than the SM Higgs self-couplings, it can be neglected to a first approximation in the 100 GeV region respect to the SM ones;
because it is of the same order in the $m_\varphi^2$ counting, it can be consistently neglected against the other beyond SM operators in dealing with TeV-scale $\varphi\varphi$ scattering.

\section{Summary and discussion \label{sec:summary}}
With the present experimental situation, the Electroweak Symmetry Breaking Sector  might be completely described by the Glashow-Weinberg-Salam Standard Model~\cite{GWS}, with 3 longitudinal $\omega_L\mid z_L$ bosons and the potential finding of its Higgs boson on the table. 
If Beyond SM physics exists, the mutual couplings of these four bosons will separate from the SM. The most interesting feature is the absence of any new particles below about 600-700 GeV implying that a separation from the SM in the couplings will lead to strong interactions. 

We have calculated and renormalized the 1-loop amplitudes in Electroweak Chiral Perturbation Theory supplemented by the new scalar boson, a natural alley of investigation based on a low-energy Effective Lagrangian that other groups are also pursuing. In doing so we have found that 3 dimension-8 derivative operators, not analyzed in depth in previous literature, are necessary in addition to the standard ones associated with $a_4$ and $a_5$. We have shown sensitivity of WBGB scattering~\cite{DHD} including now the new scalar boson, to all the LO and NLO parameters in the Lagrangian density. 

Strong interactions and unitarity violations in perturbation theory appear as soon as $v\neq f$ ($\xi\neq 1$), $\alpha\neq 1$ as seen in Eq.~(\ref{Atree}) or $\beta\neq 1$, or finally any of $\eta$, $\gamma$, $\delta$, $a_4$, $a_5\neq 0$. As usual in an effective Lagrangian, the tree-level amplitudes present polynomial behavior and the  one-loop diagrams bring in standard left and right cuts into the partial waves.

We have found that, without the NLO constants, the amplitudes behave (unsurprisingly) just like in hadron physics, with strongly attractive $I=J=0$ elastic $\omega\omega$ scattering, not so strong $I=J=1$ scattering, small (vanishing at LO) $J=2$ scattering, and repulsive $I=2$ scattering.

The latter one leaves us wondering. In QCD $\pi^+ \pi^+$ scattering is naturally repulsive due to the Pauli exclusion principle operating at the quark level, with $u\bar{d}-u\bar{d}$ blocking-off parts of the ground state wavefunctions. From the point of view of the effective Lagrangian, this repulsion is built into the flavor structure, inherited by the Electroweak Chiral Lagrangian. But it is odd, since there is no known nor necessary fermion constitution of the $W^+$ bosons, that $W^+ W^+$ ``exotic'' scattering should be repulsive. In this respect it is relieving to find, as we did in figure~\ref{wwaes}, and in agreement with~\cite{Espriu:2012ih}, that the addition of either $a_4$ or $a_5$ NLO terms with a negative sign makes the tensor amplitude much stronger at low energy. Exotic resonances are possible and leave a low-energy footstep in the EWChL (and are natural in extensions of the MSM that need a Higgs multiplet with charged members).

We wish to remark also that the $\varphi\varphi\to \varphi\varphi$ scattering occurs via a non-diagonal $\omega\omega$ loop if the interactions indeed become strong through mismatches of $\alpha$ and $\beta$ as well as $f$ and $v$ to their Standard Model values, quite irrespective of the value of the Higgs self-couplings $\lambda_3$, $\lambda_4$. These couplings do not need to be known to high precision~\cite{Gupta:2013zza} if strong interactions beyond the Standard Model operate in the TeV region, as they would have a negligible effect anyway. It appears that the Run II of the LHC should be quite conclusive as respects further strong interactions in EWSB, as long as $W_LW_L$ can be separated.

In all the amplitudes studied, the tensor $J=2$ projections computed from the LO and loop-NLO (no NLO counterterms) are totally negligible in the few-TeV regime: two-body scattering in the electroweak symmetry breaking sector is dominated by $J=0$, with a non-negligible contribution of $J=1$ in $\omega\omega\to\omega\omega$ (that because of Bose symmetry, cannot be present in the channels involving $\varphi\varphi$).

In future work we intend to study unitarization methods that render these one-loop amplitudes more theoretically sensible, and see what resonances appear in the different spin-isospin channels for physically acceptable values of the parameters.

\acknowledgments
AD thanks useful conversations with D. Espriu, M. J. Herrero and J. J. Sanz-Cillero. The work has been supported by the spanish grant FPA2011-27853-C02-01 and by the grant BES-2012-056054 (RLD).


\appendix
\section{Feynman diagrams}

In this appendix we present the Feynman diagrams that we have employed to generate the one-loop parts of the amplitudes written down in Eq.~(\ref{Aloop}), (\ref{Mloop}), (\ref{Tloop}).
The diagrams have been automatically generated with FeynRules~\cite{Alloul:2013bka} and FeynArts~\cite{Hahn:2000kx}, and evaluated with FormCalc~\cite{Hahn:1998yk,Kuipers:2012rf}, though the whole computation has also been carried out analytically, since it is of moderate difficulty, being a one-loop evaluation in the massless limit. Both ways of computing the amplitudes agree, thus checking the output of the computer programs. 
The only minor issue is that the automated tools seem to have problems with Einstein's summation convention when confronting the counterterm $(\partial_\mu \varphi \partial^\mu \varphi)^2$ that needs to be typed-in as $(\partial_\mu \varphi \partial^\mu \varphi)(\partial_\nu \varphi \partial^\nu \varphi)$. 

\begin{figure}[h] 
\begin{center}
\includegraphics*[width=\textwidth]{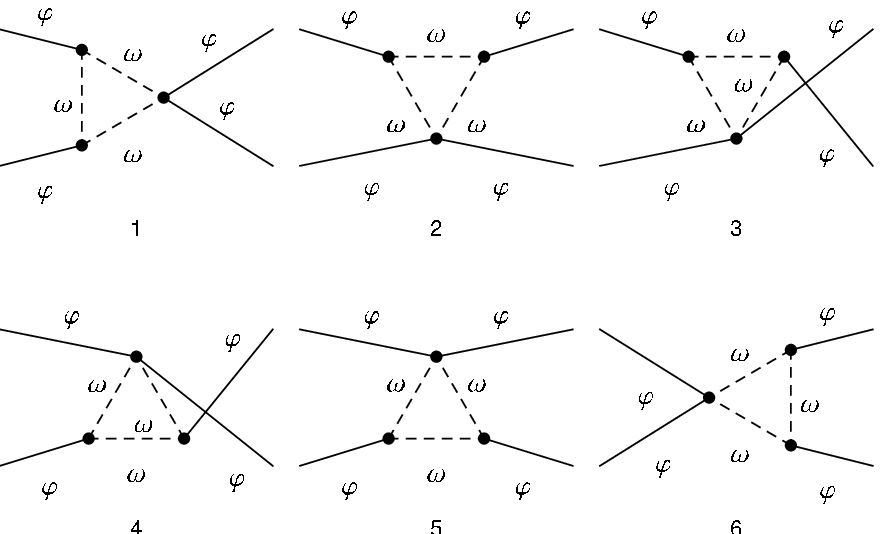}\\
\includegraphics*[width=\textwidth]{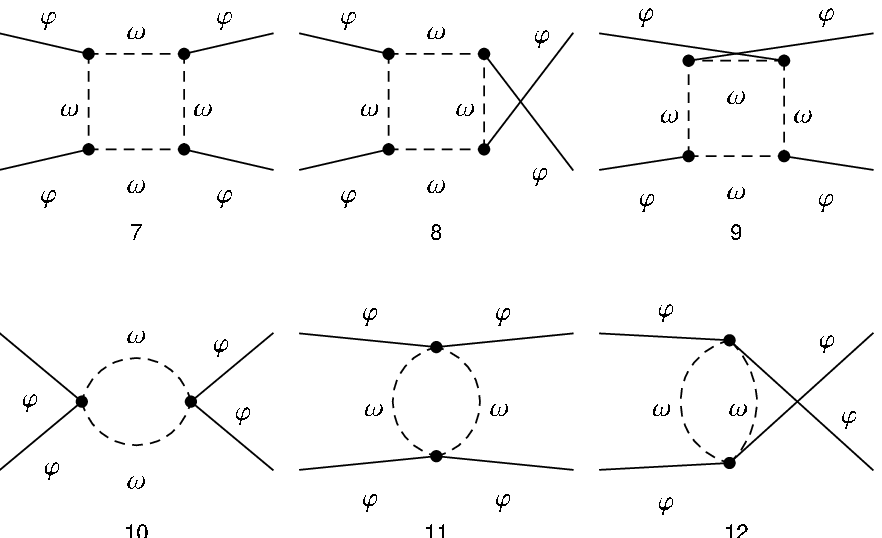}
\end{center}
\caption{\label{diags:pppp} Feynman diagrams corresponding to $\varphi\varphi$ elastic scattering.}
\end{figure}

\begin{figure}[h]
\begin{center}
\includegraphics*[width=0.6\textwidth]{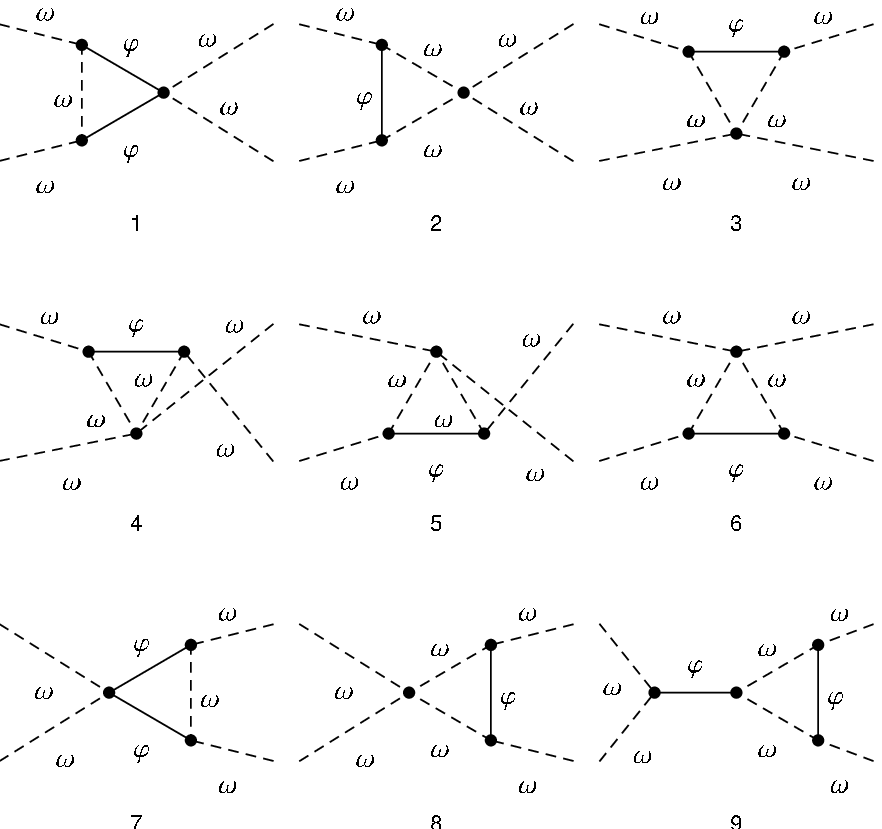}\\
\includegraphics*[width=0.6\textwidth]{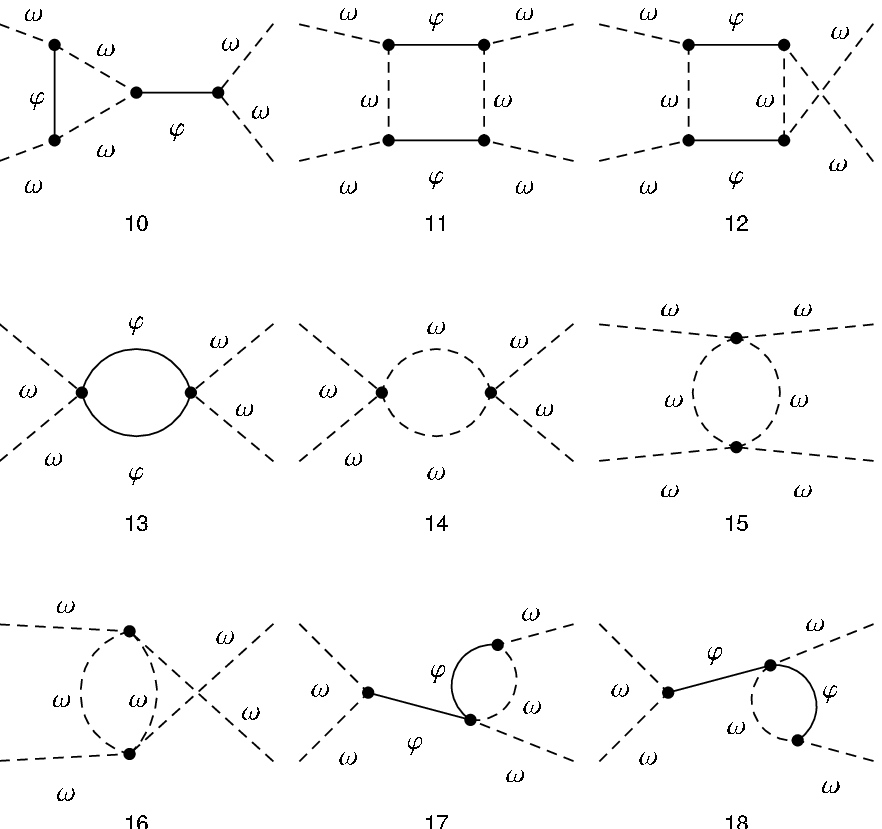}\\
\includegraphics*[width=0.6\textwidth]{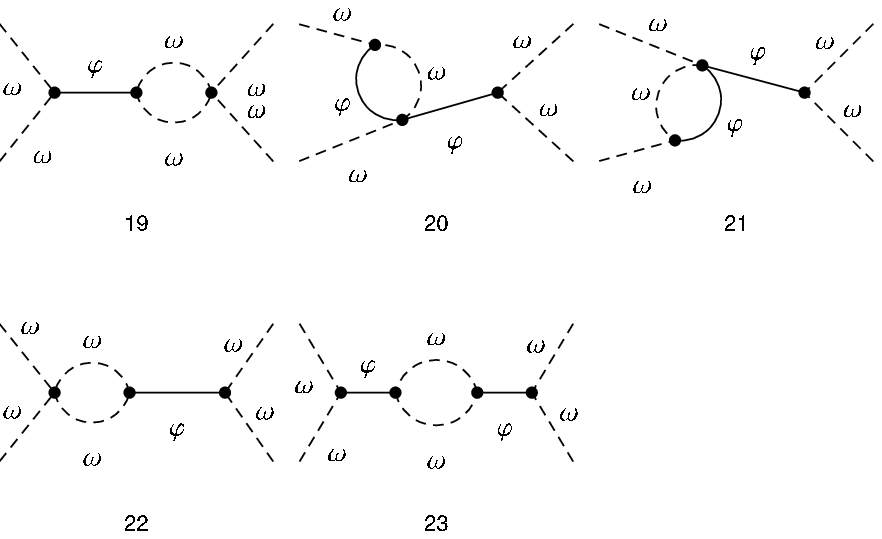}
\end{center}
\caption{\label{diags:wwww} Feynman diagrams corresponding to $\omega\omega$ elastic scattering.}
\end{figure}

Figure~(\ref{diags:wwww}) shows the $\omega\omega\to \omega\omega$ one-loop contributions that build up Eq.~(\ref{Aloop}). There one can easily identify vertex corrections (such as Feynman diagrams 1 to 10), $s$ and $t/u$ channel bubbles (diagrams 13-15) or box diagrams (11 and 12).

The same kinds of diagrams (but in smaller numbers) can be identified in figure~\ref{diags:pppp} that shows what needs to be computed for $\varphi\varphi \to \varphi\varphi$ elastic scattering in Eq.~(\ref{Tloop}).

\begin{figure}[h]
\begin{center}
\includegraphics*[height=0.45\textheight]{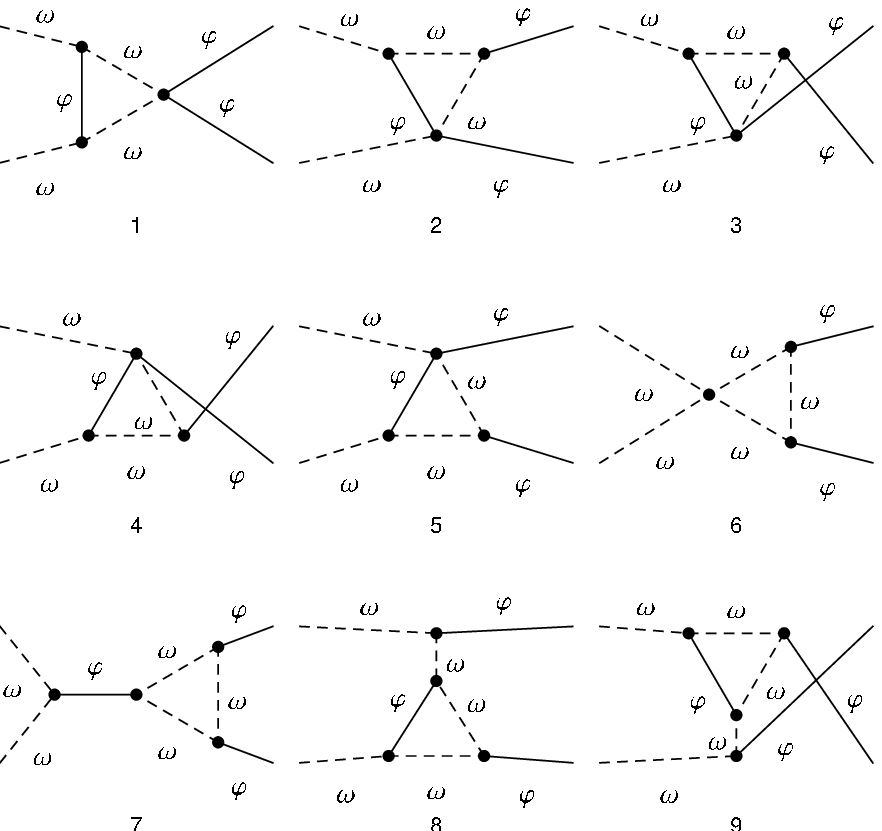}\\
\includegraphics*[height=0.45\textheight]{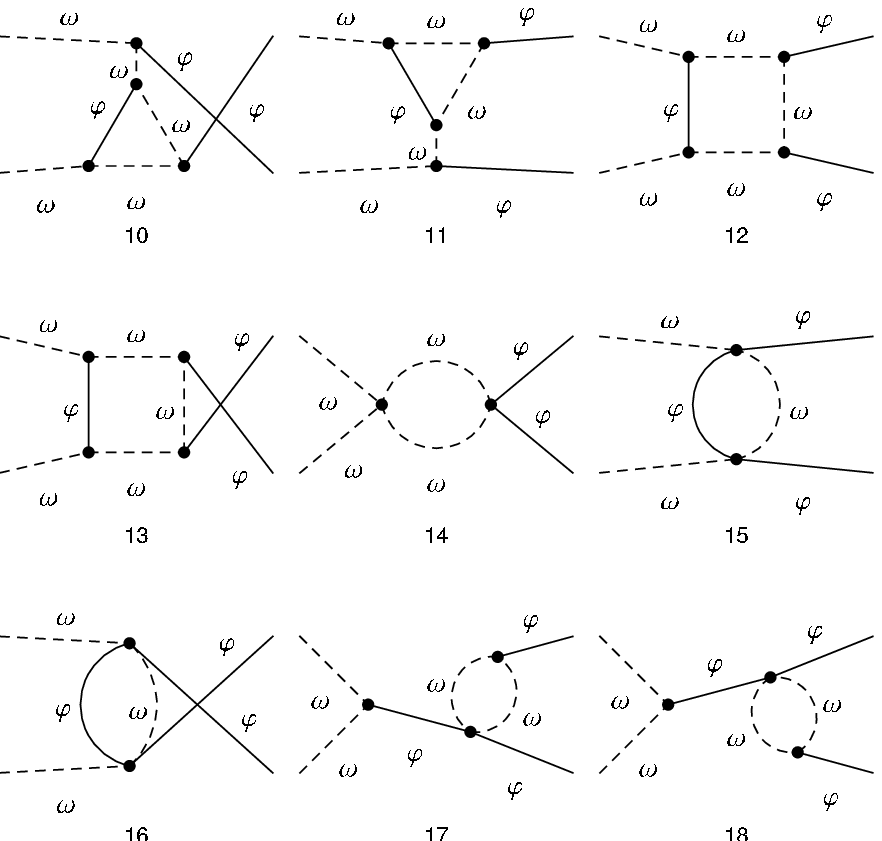}
\end{center}
\caption{\label{diags:wwppa} Feynman diagrams corresponding to $\omega\omega\to \varphi\varphi$ channel coupling.}
\end{figure}

\begin{figure}[h]
\begin{center}
\includegraphics*[width=0.8\textwidth]{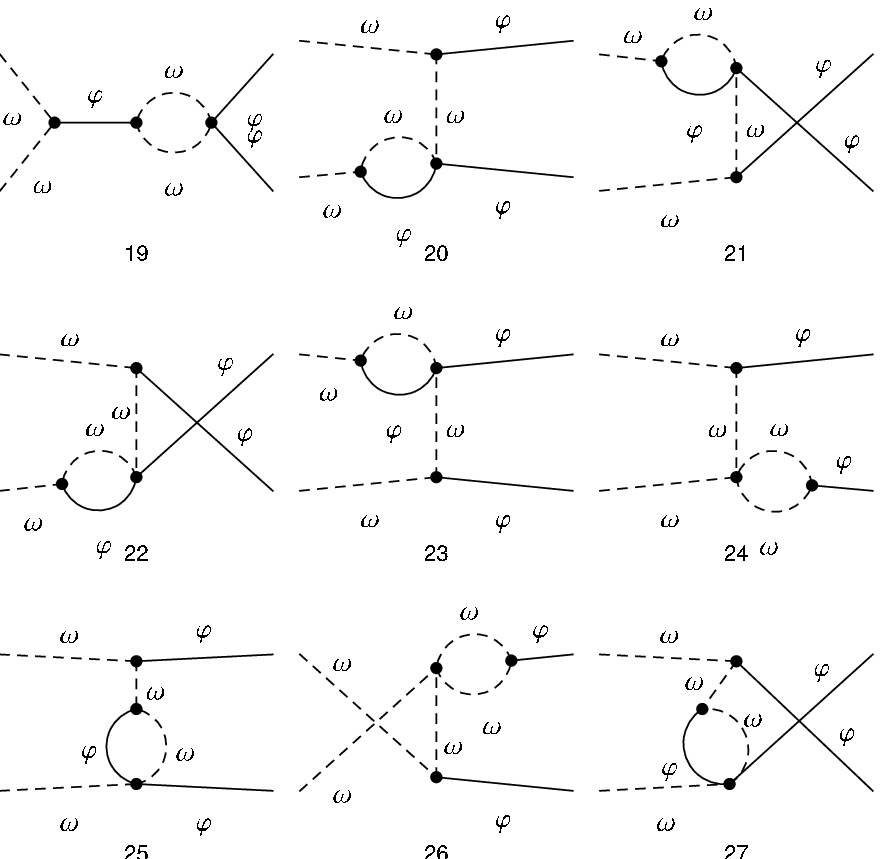}\\
\includegraphics*[width=0.8\textwidth]{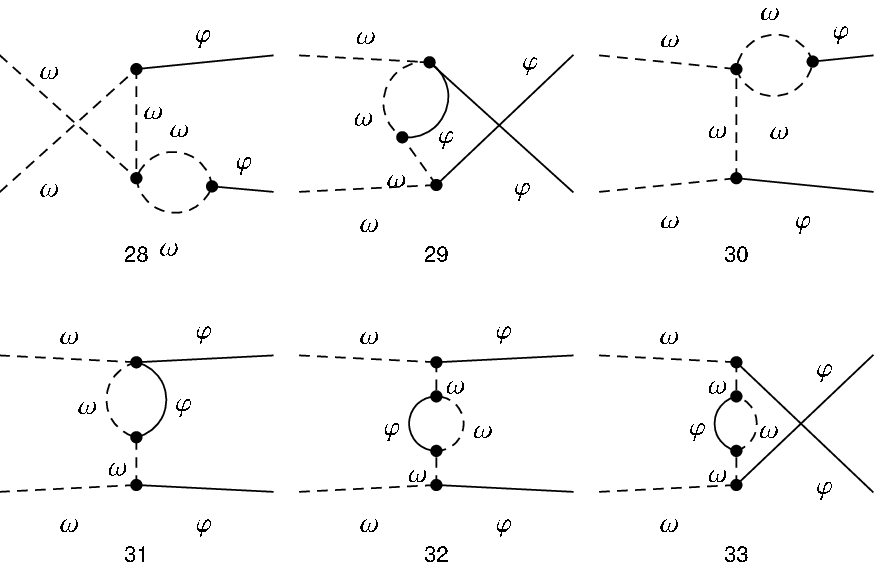}
\end{center}
\caption{\label{diags:wwppb} Further Feynman diagrams corresponding to $\omega\omega\to \varphi\varphi$ channel coupling.}
\end{figure}

Finally, figures~\ref{diags:wwppa} and~\ref{diags:wwppb} contain the Feynman diagrams necessary to compute the interchannel amplitude $\omega\omega\to \varphi\varphi$ in Eq.~(\ref{Mloop}).

\end{document}